\documentclass[12pt]{article}
\usepackage{amsmath}
\usepackage{amssymb}
\usepackage{epsfig}
\usepackage{cite}
\usepackage{color,colordvi}

\newcommand{\eq}{\begin{equation}}
\newcommand{\eqx}{\end{equation}}
\newcommand{\eqn}{\begin{eqnarray}}
\newcommand{\bi}{\begin{itemize}}
\newcommand{\eqnx}{\end{eqnarray}}
\newcommand{\ei}{\end{itemize}}

\newcounter{hran}

\def\MSbar{\relax\ifmmode\overline{\rm MS}\else{$\overline{\rm MS}${ }}\fi}

\begin{document}

\begin{center}

{\Large \textbf{Probing Quantum Geometry at LHC}}

\vspace{1cm}

\textbf{Gia Dvali}$^{a,b,d,c}$, \textbf{Cesar Gomez}$^{e}$ and \textbf{Slava
Mukhanov}$^{a,b,d}$

\vspace{.6truecm}

\vspace{.2truecm}

\emph{$^a$Arnold Sommerfeld Center for Theoretical Physics\\[0pt]
Department f\"ur Physik, Ludwig-Maximilians-Universit\"at M\"unchen\\[0pt]
Theresienstr.~37, 80333 M\"unchen, Germany}


\emph{$^b$Max-Planck-Institut f\"ur Physik\\[0pt]
F\"ohringer Ring 6, 80805 M\"unchen, Germany}


\emph{$^c$CERN, Theory Division\\[0pt]
1211 Geneva 23, Switzerland}


\emph{$^d$CCPP, Department of Physics, New York University\\[0pt]
4 Washington Place, New York, NY 10003, USA}

\vspace{.2truecm}

\emph{$^e$ Instituto de F\'{\i}sica Te\'orica UAM-CSIC, C-XVI \\[0pt]
Universidad Aut\'onoma de Madrid, Cantoblanco, 28049 Madrid, Spain}\\[0pt]
\end{center}


\centerline{\bf Abstract} 
We present  an evidence, that the volumes of compactified spaces as well as
the areas of black hole horizons must be quantized in Planck units. This
quantization has phenomenological consequences, most dramatic being for
micro black holes in the theories with TeV scale gravity that can be
produced at LHC. We predict that black holes come in form of a discrete
tower with well defined spacing. Instead of thermal evaporation, they decay
through the sequence of spontaneous particle emissions, with each transition
reducing the horizon area by strictly integer number of  Planck units. Quantization of the
horizons can be a crucial missing link by which the notion of the minimal
length in gravity eliminates physical singularities. In case when the
remnants of the black holes with the minimal possible area and mass of order
few TeV  are stable, they might be good candidates for the cold dark
matter in the Universe.

\vskip .4in \noindent


\section{Introduction}

The black hole physics dictates the existence of the minimal possible
measurable scales for the area and the volume. In fact, in any experiment,
which is supposed to measure area (cross section) smaller than the Planck
area $L_{Pl}^{2}\simeq 10^{-66}cm^{2}$ one has to involve particle scattering with the
higher than Planckian momentum-transfer, which will form a black hole with
the radius exceeding the Planck length $L_{Pl}.$ Impossibility to probe
sub-Planckian length-scales suggests that Einstein gravity viewed as field
theory is ultra-violet self-complete, but in a sense profoundly different
from the conventional (Wilsonian) notion\cite{gia-cesar1}. The key
ingredient is a crossover between the quantum degrees of freedom and
classical states (Black holes).  This phenomenon among other things implies, that the
smallest black holes, with mass $\sim \, L_{Pl}^{-1}$, become sharp quantum
resonances, which instead of conventional thermal evaporation are either
stable or decay into the light fields in a single quantum jump.

These two facts, impossibility of resolving distances shorter than $L_{PL}$
and crossover of classical black holes into the quantum resonances, suggest
that quantum jumps should not be property of exclusively smallest black
holes. Rather, black holes must obey some fundamental quantization rule
which for a large black hole should reproduce the conventional thermal
picture, but should  simultaneously  explain the gradual transition of the smaller and smaller  black
holes into the quantum particles.

We shall argue, that the right quantization condition is that the area must
be quantized in terms of the integers of the Planck area. In particular,
this refers to the well defined area of the black hole horizon. The above
assumption is well supported by the fact that the black hole horizon area
behaves as adiabatic invariant \cite{Bek1} and by statistical interpretation
of the entropy of the quantized black hole \cite{Muk1}. 

Moreover, we wish to show,  that from the
requirement that black hole saturate the holographic bound on the
information storage\cite{holog} it follows that in any elementary emission its horizon
area cannot be reduced by less than one Planck unit. In fact, let us arrange
a \textquotedblleft gedanken experiment\textquotedblright\ in which we tag
the information bit without altering the classical black hole geometry. One
can achieve this as follows. Assume that we have a spectator light particle
that carries a charge under a gauged $Z_{2}$-symmetry, and consider black
hole evaporation in such a theory. For example, one can prepare a large
classical black hole and endow it with a $Z_{2}$-charge by throwing a $Z_{2}$%
-charged particle in it. Since $Z_{2}$-charge is not associated with any
massless electric field, there exist no classical hair under it \cite{nohair}, and the effects
of quantum hair \cite{quantumhair}  on the classical geometry can be safely ignored. The black
hole will then evaporate as the usual neutral Schwarzschild black hole would
do. Now consider an elementary act of emission of $Z_{2}$-charged particle.
Due to the absence of hair, emission of this particle is not any different
from the emission of any other quantum, but the advantage is that the $Z_{2}$%
-charge allows us to trace the information in a clean way. There is a full
single bit of information encoded in this charge. Therefore it is obvious
that after particle emission, the horizon area cannot shrink by less than
one Planck unit, because otherwise the remaining surface would violate the
requirement of the saturation of holographic bound. Likewise, we could
arrive to an inconsistency in the inverse process of absorption. \footnote{ Had we instead used either a single particle with a $Z_N$-charge or a single member of 
$N$ distinct particle species family, the information would change by $N$ bits, 
changing the black hole area by $NL_{Pl}^2$. This  fully agrees with the earlier findings \cite{bound, zn, gia-cesar}, that in theories with $Z_N$-symmetry or  $N$-distinct particle species  the fundamental length is $\sqrt{N}L_{Pl}$, and the area of {\it small}  black holes  (for which  gravity 
is modified )  could instead  be  quantized in these units.  However,  due to modification  of mass-to-radius  relation this gives exactly the same result as quantization of the effective area
measured by massless graviton in units of $L_{Pl}^2$! 
  Below we shall witness this connection on explicit example of Kaluza-Klein theories, which propagate tower of modes.}   

 Thus, we
see that the only way to maintain holographic bound to be saturated is that
elementary jump of the horizon area in any one-particle emission process
must be quantized in Planck units, in accordance with the conjecture of \cite%
{Bek1}, \cite{Muk1}, \cite{BM}. The essence of our argument is to  transform the abstract concept 
of information into  a measurable $Z_2$-charge with no classical-hair.

The area quantization can have a dramatic consequences for the Hawking
evaporation of the black holes, especially when it refers to the
microscopic black holes \cite{Muk1}, \cite{BM}. The quantization of the
geometry of the black hole horizon become especially interesting in the
theories with large extra dimensions, where the fundamental Planck scale can be lowered by many
orders of magnitude and become accessible at LHC.

\section{Quantum Geometry in Gravity with extra dimensions}

In the theories with extra-dimensions gravity can become $D=4+d$ dimensional
on submillimeter or smaller scales and the corresponding quantum gravity
Planck scale can be lowed to the scale reachable at  LHC  \cite{ADD, AADD}. In
fact, assuming that the fundamental theory is a $D$-dimensional Einstein
gravity, let us consider compactification of $D$-dimensional space-time to
the direct product of $4$-dimensional Minkowski space-time and of a $d$%
-dimensional compact torus of volume $V_{d}$. Then, it follows from\footnote{%
To simplify the formulae we are skipping the numerical coefficients such as $4\pi$
etc., which reader can easily restore.} 
\begin{equation}
\frac{1}{L_{D}^{2+d}}\int R\sqrt{-g}d^{4+d}x=\frac{V_{d}}{L_{D}^{2+d}}\int R%
\sqrt{-g}d^{4}x=\frac{1}{L_{Pl}^{2}}\int R\sqrt{-g}d^{4}x  \label{1} \, 
\end{equation}%
that the corresponding 4-dimensional and $D$- dimensional Planck lengths, denoted by
$L_{Pl}$ and $L_{D}$  respectively, are related as%
\begin{equation}
L_{D}\,=\,(V_{d}/L_{D}^{d})^{{\frac{1}{2}}}\,L_{Pl}  \label{2} \, .
\end{equation}%
The key point is, that the ratio 
\begin{equation}
(V_{d}/L_{D}^{d})\,=\,N\,,  \label{quant}
\end{equation}%
counts the total number of Kaluza-Klein modes with masses not exceeding $%
L_{D}^{-1}$ \cite{bound, gia-cesar}.    More massive species should not be counted 
since they form black holes. With this assumption we arrive to the conclusion
that the volume of the compact torus is quantized in the fundamental Planck
units. 

 
  Let us illustrate the point on a simple case of $D=5$. Let  the compactification volume be  $V_1 \, \equiv \, R$ and let us assume $R/L_5 \,  = \,  N + \epsilon$,  where  $\epsilon <1$. If this were possible,  then we could choose  $N=0$,  
 meaning that $R=\epsilon L_5$.  That is, we could take  the compactification radius arbitrarily smaller than $L_5$.  But in this case the fifth dimension stops existence in any physical sense.  In particular,  $L_5$ becomes shorter than the four-dimensional  Planck length, $L_{Pl}$,  and 
 the would-be  KK modes  become ultra-Planckian mass states and are indistinguishable from four-dimensional classical black holes,  according to \cite{gia-cesar1}.   The extra dimension ceases 
 to exist.  In other words,  quantization of compact volume automatically follows if $R$ is forced to be bigger than $L_5$ and this should be the case due to  the fact that trans-Planckian states are equivalent to classical black holes.

For $N\simeq 10^{32}$ the $D$-dimensional Planck length $L_{D}$
becomes of order $L_{D}\simeq 10^{-17}cm$ and corresponds to $D$%
-dimensional nonperturbative quantum gravity scale reachable at LHC.
Moreover, from the relation between the mass and gravitational radius, 
\begin{equation*}
M_{BH}=\frac{R_{g}^{1+d}}{L_{D}^{2+d}},
\end{equation*}%
it follows that the minimal black hole with $R_{g}\simeq L_{D}$ has mass $%
M_{BH}\simeq L_{D}^{-1}\simeq O\left( 1\right) Tev$ and hence such black
holes can be produced at LHC. They should be described by a nonperturbative $%
D-$dimensional quantum gravity which can substantially revise the process of
their Hawking evaporation. In fact, below we will show that the assumption
about the area quantization leads to a dramatic modification of the
Hawking's result, obtained in the approximation of the external classical
gravitational field.

In $D$-dimensional quantum gravity the area of the $D$-dimensional
Schwarzschild black hole should satisfy the following quantization rule 
\begin{equation}
A_{D}\,=\,n\,L_{D}^{D-2}\,.  \label{quantization}
\end{equation}%
This means, that an effective projection of the horizon area $A_{D},$
measured by a $D$-dimensional observer$\,$, must be equal to the integer
number of the corresponding $D$-dimensional Planck units.

One may wonder in which units the area should be quantized from the point of
view of $4-$dimensional observer when $d=D-4$ dimensions are compactified on
the torus of volume $V_{d}\,\equiv \,R^{d}$, where $R$ is the
compactification radius. Let us show that in this case the quantization rule
(\ref{quantization}) implies that the $4$-dimensional projection of the the
black hole area is automatically quantized in terms of 4d Planck units $L_{Pl}^{2}\simeq
10^{-66}cm^{2}.$  This is a manifestation of a general phenomenon that the 
Bekenstein-Hawking entropies computed for one and the same black hole by 
$D$-dimensional and $D-d$-dimensional observers in their respective Planck units, exactly 
match\cite{solod}.

We consider first a black hole of radius $R_{g}\,\gg \,R$.
The horizon of such black hole has a geometry of $S_{2}\,\times \,T_{d}$,
where $S_{2}$ is an usual $4$-dimensional Schwarzschild two-sphere of area $%
A_{4}\,=\,R_{g}^{2}$, and $T_{d}$ is a $d$-dimensional torus of volume $%
V_{d}\,=\,R^{d}$; hence, $A_{D}\,=\,A_{4}\,V_{d}$. \ According to (\ref%
{quantization}) the $2+d$-dimensional horizon area should be quantized in
units of $L_{D}^{2+d}$ and we obtain 
\begin{equation}
n\,=\,A_{D}\,/L_{D}^{2+d}\,=\,R_{g}^{2}V_{d}/L_{D}^{2+d}\,=A_{4}/L_{Pl}^{2},
\label{Darea}
\end{equation}%
where we have taken into account relation (\ref{2}). Thus, we have shown, that  the
projection of the horizon of a $D$-dimensional black hole in 4$d$ space-time
consists of as many 4$d$ Planck areas $L_{Pl}^{2}$ as the original $D-2$
dimensional horizon contains corresponding Planck units $L_{D}^{2+d}$.

For the black hole of size $R_{g}\,\ll \,R$, the matching of two
quantizations conditions is a little bit more subtle. The geometry of such a
black hole to a very good approximation is given by  $D$-dimensional Schwarzschild
geometry, with the horizon being a ${2+d}$-dimensional sphere of area $%
A_{D}\,=\,R_{g}^{2+d}$. However, a four-dimensional observer only measures a 
$4$-dimensional projection of this area, which is a two-dimensional sphere
of the effective area $A_{4}\,=\,R_{g}^{2}(R_{g}^{d}/V_{d})$. The additional
factor $(R_{g}^{d}/V_{d})$ is due to the suppression of the overlap of the
KK wave-function profile in $T_{d}$ with the area of the BH. Quantizing this
area in units of four-dimensional Planck area $L_{Pl}^{2}$, we get, 
\begin{equation}
n\,=\,A_{4}/L_{Pl}^{2}\,=\,\,R_{g}^{2}\,(R_{g}^{d}/V_{d})\,L_{Pl}^{-2}\,,
\label{4areasmall}
\end{equation}%
or, using the relation (\ref{2}), we can rewrite this as, 
\begin{equation}
n\,=\,A_{D}\,/L_{D}^{2+d}\,.  \label{Dareasmall}
\end{equation}%
The latter expression is a quantization of the area in units $L_{D}^{2+d}$
with exactly the same $n$!

This is a remarkable result, which means that the entropy of the quantized
black hole, which according to \cite{Muk1} is equal to 
\begin{equation*}
S=\left( n-1\right) \ln 2,
\end{equation*}%
is the same from the points of view of $D$-dimensional and $4$-dimensional observers and
it is proportional to the respective areas measured in the corresponding Planck units.

\section{Predictions}

There are two obvious places where one could look for the experimental
signatures of our results. These are micro black holes, that may be
accessible at LHC, and cosmology.

Not surprisingly, most dramatic phenomenological impact the horizon area quantization has on
the microscopic black holes of about $TeV$ mass, which are predicted in the
scenario with large extra dimensions\cite{ADD, AADD}. Let us focus on the case
when the size of the black holes, that can be potentially produced at LHC,
is much smaller than the compactification radius and hence the effects of
compactification can be ignored.

The mass of the Schwarzschild black hole in $D=4+d$ dimensions can be
written in terms of the horizon area $A$ as, 
\begin{equation}
M\,=\,M_{D}^{2+d}\,A^{{\frac{1+d}{2+d}}}\ ,  \label{BHmass}
\end{equation}%
where $M_{D}=L_{D}^{-1}$ is the minimal possible mass for the black hole.
The quantization condition (\ref{quantization}) implies that the mass is
also quantized,%
\begin{equation}
M_{n}=M_{D}n^{{\frac{1+d}{2+d}}} \, , \label{massquantization} 
\end{equation}%
and if the distance between levels is larger than the width of the levels
the black hole evaporation has to be considered as the sequence of discrete
transitions. The elementary transition between two nearby levels corresponds
to the change of the area by one Plank unit $L_{D}^{d+2}.$ As a result of
such transition the black hole emits a particle of the energy 
\begin{equation}
\varepsilon _{n}=M_{n}-M_{n-1}\,=\,M_{D}\left( n^{{\frac{1+d}{2+d}}%
}\,-\,(n-1)^{{\frac{1+d}{2+d}}}\right) \,.  \label{jump}
\end{equation}%
For $n\,\gg \,1$ this can be approximated by 
\begin{equation}
\varepsilon _{n}\,\simeq \,{\frac{1+d}{2+d}}\,R_{g}^{-1}\,.  \label{jump1}
\end{equation}%
For a large black hole, with $R_{g}\,\gg \,L_{D}$, the direct transitions
from level $n$ to $n-2$ and other nearby levels reproduce the thermal
properties of spectrum of the radiation emitted from the \textquotedblleft
box\textquotedblright\ of finite size $R_{g}$ with an effective Hawking
temperature $T\,\equiv \,R_{g}^{-1}$ \cite{BM}. However, the crucial
difference is that the large wavelengths exceeding $R_{g}$ are not present
for an arbitrarily large black hole. For small black holes, that are of LHC
interest, the difference is even more dramatic. Rather than viewed as
evaporation, the decay process of a small black hole represents a sequence
of spontaneous emissions with the total number of emitted quanta given by
\begin{equation}
n_{total}\,\simeq \,(M_{BH}L_{D})^{{\frac{2+d}{1+d}}}\,.  \label{total}
\end{equation}%
We shall now estimate the emission rate and the lifetime of the black
without relying on the thermal properties of radiation.

Let us consider the decay via graviton emission, treating it as a sequence
of the spontaneous emissions. Since each step is a quantum emission process
with a momentum transfer $\varepsilon _{n}$, by general covariance \cite{dvali-lust} the
graviton emission rate per unit of the Planck area, corresponding to such
momentum-transfer is, 
\begin{equation}
\tilde{\Gamma}\,_{n}=\alpha \,(L_{D}\varepsilon _{n})^{2+d}\,\varepsilon
_{n}\,=\alpha M_{D}\,\left( n^{{\frac{1+d}{2+d}}}\,-\,(n-1)^{{\frac{1+d}{2+d}%
}}\right) ^{3+d},  \label{rategraviton}
\end{equation}%
where $\alpha $ is some numerical coefficient which seems to be
significantly smaller than unity because of the existence of the external
classical barrier \cite{BM} \footnote{%
Note that due to smallness of $\alpha $ the smallest black holes are sharp quantum
resonances in accordance with findings of \cite{gia-cesar}}. To get the
total emission rate we have to multiply (\ref{rategraviton}) by the areas in
the Planck units and hence the total emission rate is%
\begin{equation}
\Gamma _{n}=\tilde{\Gamma}\,_{n}\left( \frac{R_{g}^{2+d}}{L_{D}^{2+d}}%
\right) =n\tilde{\Gamma}\,_{n}.  \label{6}
\end{equation}%
This expression describes the rate of transition $n\,\rightarrow \,n-1.$
Taking into account (\ref{jump1}) for large $n$ it can be approximated as%
\begin{equation}
\Gamma _{n}\simeq \alpha \,\,\varepsilon _{n}  \label{5}
\end{equation}%
Because $\alpha $ seems to be much smaller than unity even with taking into
account the emission of other particles \cite{BM} the width of the levels,
proportional to $\Gamma _{n},$ is much smaller than the distance between the
levels $\varepsilon _{n}$. Therefore one expects that small black hole will
emit particles in rather sharp lines corresponding to the energy difference $%
\varepsilon _{n}$. The characteristic time of the transition with one
particle emission is given by, 
\begin{equation}
\tau _{n}\,\simeq \,\Gamma _{n}^{-1}\,=n^{-1}\tilde{\Gamma}_{n}^{-1}\,.
\label{lifen}
\end{equation}%
For a large black hole, $n\,\gg \,1,$ the half life-time,
corresponding to the emission of about $n$ quanta, obtained using
approximations (\ref{jump1}), (\ref{5}), is 
\begin{equation}
T_{n}\simeq n\tau _{n}\,\sim \,\alpha ^{-1}n^{{\frac{3+d}{2+d}}%
}\,L_{D}\,\sim \alpha ^{-1}R_{g}^{3+d}\,M_{D}^{2+d}\,.  \label{lifelarge}
\end{equation}%

For  small black holes one has to use more precise expressions (\ref{jump}), (%
\ref{rategraviton}) and (\ref{6}) instead. These expressions indicate that
a small black hole essentially decays as a quantum particle. This is in
agreement with the finding of \cite{gia-cesar1}, that Einstein gravity must propagate
quantum degrees of freedom with mass of order $M_{D},$ which can be viewed
as the very end points of black hole evaporation. The above results fully
support this picture. Thus as anticipated, even if LHC reaches the threshold
of the black formation, the first accessible quantum-gravity states will be
indistinguishable from the particles.

The maximal values of $n$ that can be probed at center of mass energy $E_{c}$%
, is given by $n\,=\,(E_{c}/M_{D})^{{\frac{d+2}{d+1}}}$, and the momenta of
the first decay products soften according to the following law, 
\begin{equation}
p\,=\,M_{D}\left( (E_{c}/M_{D})\,-\,((E_{c}/M_{D})^{{\frac{d+2}{d+1}}}-1)^{{%
\frac{1+d}{2+d}}}\right) \,.  \label{soft}
\end{equation}%
This softening is the characteristic signal of the beginning of the black
hole production at LHC, but its detection will requires reaching higher  values of $n$. 

\section{Conclusions and Speculations}

We have argued that the existence of the minimal length in gravity, when
combined with black hole no-hair properties and holographic bound on information
storage,  implies quantization of geometric areas in Planck units. We have
discussed some fundamental and phenomenological implications of this
quantization. Below we shall briefly summarize these implications as well as some
future prospects for the quantized geometry.

\textbf{Black Hole Tower at LHC. }A most dramatic impact quantization of the
black hole horizon has for TeV mass black holes that are predicted in
theories with low Planck mass scenarios \cite{ADD, AADD, bound} \footnote{%
Note by V.M. : if it is confirmed then Gia will buy me 10$^{3}$ bottles of
wine, as opposed to $10^{32}$ particles that these theories propagate.} . As
shown above, in pure gravity such black holes will come in form of the
quantum resonances spaced according to (\ref{jump}). In realistic scenario
one also has to include Standard Model particle localized on a 3-brane. As
shown in \cite{gia-cesar}, their presence increases the minimal length to $%
L_{N}\,=\,N_{SM}^{{\frac{1}{d+2}}}L_{D}$, where $N_{SM}$ is the number of
Standard model helicities per helicity of a $4+d$-dimensional graviton. The
black hole horizon now must be quantized in $L_{N}$-units, but due to a
small number of Standard Model species and the suppression by the
exponential factor $1/(d+2)$, sensitivity in the change is inessential, 
\begin{equation*}
M_{n}=\left( \frac{M_{D}}{N_{SM}^{{\frac{1}{d+2}}}}\right) \,n^{{\frac{1+d}{%
2+d}}}\,.
\end{equation*}%
Already for $d=2$ this renormalizes the quantization unit only by factor of $
2$. So the influence of the Standard Model  particles on the black hole quantization can
be ignored.

The above quantization implies that the properties of the micro black holes
are nothing like the conventional view of semi-classical thermal states.
Instead, they will come as sharp quantum resonances, with characteristic
trajectory given by (\ref{massquantization}).

Micro black holes can have  essentially any quantum number that can be
composed out of particle states in the Standard Model, and thus, can imitate
many weakly coupled extensions of the Standard Model, such as, extra 
gauge-bosons, supersymmetry, or other exotic models. Thus, in certain sense gravity
projects the spectrum of all possible extensions of the Standard
Model at the fixed scale around $M_{D}$. The only way to recognize the black
hole nature of these new states, is to create highly excited black holes  that should
obey the spacing (\ref{jump}).

In particular, if, as suggested in \cite{gia-cesar}, the scattering of
longitudinal $W^{\pm}$-bosons gets unitarized by black holes, the ``fake"
Higgs particle will be a lowest excitation of one such tower.

The interesting phenomenological question would be to differentiate the
black hole tower from a more conventional string Regge trajectory, also
predicted in \cite{AADD} (for  more detailed phenomenological  studies of this effect see, \cite{peskin, dieter, tao}). Such a tower, will open up first, if the Standard Model is
UV-completed by a weakly-coupled string theory, but will be replaced by the
black hole tower if UV-completion happens by gravity (or if string theory is
strongly coupled).

\textbf{\ Implication for Cosmological Singularities. } The area-quantization can
be a crucial missing link in reconciling the notion of the minimal length,
that is responsible for self-completeness of gravity, with cosmological
singularities.

Indeed, if the Hubble volume is quantized in $L_{Pl}$-units, the final states
of the collapsing universe may be similar to the black hole
evaporation. As we have argued above, a black hole with $n=1$ will decay into
a single quantum jump into the lighter states. Similarly, the collapsing
Universe after reaching the  Planckian density will represent collection 
of Hubble patches each of the area $n=1$. Despite the fact that number of patches 
may be enormous, each patch can only undergo a discrete quantum transition which changes $n$ 
by an integer.   A transition from  $n = 1$ to $n=0$ corresponds to the decay of a given patch into 
nothingness.  Since the patches are causally disconnected,  we expect that  the transitions are 
not correlated.  In this way, the total volume of the Universe is reduced by  an 
integer Planck volumes, without further increase of density. 
  The Hubble patches with $n=1$ decay into nothingness
within a single  quantum jump, without passing through singularity.

Similarly, Universe can be created out of nothing as a
result of a quantum transition \cite{quantumcreation}, but  since $n$ can change only by an integer
in any Hubble patch, passing through a classical singularity  is avoided. 
This concepts require further study that
will not be attempted here.

\textbf{Implication for Moduli.} 
Quantization of compactification volume implies quantization on the vacuum expectation values 
of the moduli fields.  In this way moduli cannot assume arbitrary values, and this may shed some 
new light on issues of moduli stabilization and perhaps on supersymmetry breaking. 
Consequently,  the values of the coupling constants  that are set by moduli fields should also obey 
certain quantization rule. 
In the same spirit, applying quantization of areas in Planck units to string theory  and taking into the account  the fact that string length must play the role of the fundamental scale,  we conclude that 
string coupling,  $1/g_s^2$, should also be quantized.  This  is in agreement with the results 
of \cite{dvali-lust, gia-cesar2} according to which  $1/g_s^2$ measures the effective number of particles  species to which the semi-classical black holes can decay normalized to the rate of 
emission of graviton helicities.

\textbf{Implication for Time-Dependent Backgrounds. }The other important
application of our results is for cosmological time-dependent backgrounds,
in which horizons and volumes change. In particular, it follows that change
of expectation values of the scalar fields in the early universe cannot be
continues, but rather has to proceed via quantum jumps, each of which
changes compactifiction volume by one Planck unit. This discretization of
motion is especially important for small compactification volumes, and could
have imprint in early cosmology, especially when one considers the self
reproducing eternal universe.


\textbf{Implications for Cold Dark Matter. }The geometry quantization may
also have an interesting cosmological applications for Dark Matter. It might
happen that the minimal black holes of mass $M_{D}\simeq O(1)Tev$
(corresponding to $n=1$) are stable. Then they have about right geometrical
cross section to be good candidate for the cold dark matter in our universe.




  \vspace{5mm}
\centerline{\bf Acknowledgments}

The work of G.D. was supported in part by Humboldt Foundation under Alexander von Humboldt Professorship,  by European Commission  under 
the ERC advanced grant 226371,  by  David and Lucile  Packard Foundation Fellowship for  Science and Engineering and  by the NSF grant PHY-0758032. 
The work of C.G. was supported in part by Grants: FPA 2009-07908, CPAN (CSD2007-00042) and HEPHACOS P-ESP00346.
V.M. is supported by TRR 33 \textquotedblleft The Dark
Universe\textquotedblright\ and the Cluster of Excellence EXC 153
\textquotedblleft Origin and Structure of the Universe\textquotedblright .



\begin{thebibliography}{9}

\bibitem{gia-cesar1} 

G.~Dvali and C.~Gomez, Self-Completeness of Einstein Gravity, 
arXiv:1005.3497 [hep-th]; 

G.~Dvali, S.~Folkerts, C.~Germani,  Physics of Trans-Planckian Gravity, 
arXiv:1006.0984 [hep-th]. 


\bibitem{Bek1} J. D. Bekenstein, Lett. Nuovo Cimento 11, 467 (1974).

\bibitem{Muk1} V. Mukhanov, Pis. Eksp. Teor. Fiz. 44, 50 (1986) [JETP
Letters 44, 63 (1986)], and in Complexity, Entropy and the Physics of
Information, SFI Studies in the Sciences of Complexity, vol. III, ed. W. H.
Zurek (Addison--Wesley, New York 1990).

\bibitem{BM} J. D. Bekenstein, V. Mukhanov, Phys.Lett. B360, 7 (1995). 


\bibitem{holog} 

J. D. Bekenstein,  Phys. Rev. D 23, 287 (1981); 

 W. G. Unruh and R. M. Wald, Phys. Rev. D 25, 942 (1982); 27, 2271 (1983)
 
 J. D. Bekenstein, Phys. Rev. D 26, 950 (1982);  Phys.
Rev. D 49, 1912 (1994). 

G.~'t Hooft,
"Dimensional reduction in quantum gravity", gr-qc/9310026.

L.~Susskind,
"The World As A Hologram", J.\ Math.\ Phys.\  {\bf 36}, 6377 (1995),  hep-th/9409089.

L.~ Susskind and  E.~ Witten, ``The Holographic bound in anti-de Sitter space", 
 arXiv:9805114 [hep-th]. 

\bibitem{nohair} 

  W.~Israel,  {\it Phys. Rev.} {\bf 164} (1967) 1776;  {\it Commun. Math. Phys.}
{\bf 8}, (1968) 245;  

B.~ Carter, {\it Phys. Rev. Lett.}  {\bf 26}  (1971)  331.

J.~Hartle, {\it Phys. Rev.}  {\bf D 3} (1971) 2938.

  J.~Bekenstein, {\it Phys. Rev. }\  {\bf D 5}, 1239 (1972);
  {\it Phys.\ Rev.}\  {\bf  D 5},  (1972) 2403; {\it Phys. Rev. Lett.} {\bf 28} (1972) 452. 
  
  C.~Teitelboim, {\it Phys. Rev.} {\bf D 5} (1972) 294.  
  
  
 \bibitem{quantumhair}

   L. M. Krauss and F. Wilczek, Phys. Rev. Lett. 62, 1221 (1989) ; 
   
   S. R. Coleman, J. Preskill and F. Wilczek, Mod. Phys. Lett. A 6, 1631 (1991);
Phys. Rev. Lett. 67, 1975 (1991); Nucl. Phys. B 378, 175 (1992) [arXiv:hep-th/9201059].
  
  
   
\bibitem{ADD}
  N.~Arkani-Hamed, S.~Dimopoulos and G.~R.~Dvali,
  ``The hierarchy problem and new dimensions at a millimeter,''
  Phys.\ Lett.\  B {\bf 429}, 263 (1998)
  [arXiv:hep-ph/9803315];

\bibitem{AADD}
 I.~Antoniadis, N.~Arkani-Hamed, S.~Dimopoulos and G.~R.~Dvali,
  ``New dimensions at a millimeter to a Fermi and superstrings at a TeV,''
  Phys.\ Lett.\  B {\bf 436}, 257 (1998)
  [arXiv:hep-ph/9804398].






\bibitem{bound}

G.~Dvali, ``Black Holes and Large N Species Solution to the
Hierarchy Problem,'' arXiv:0706.2050 [hep-th];

G.~Dvali and M.~Redi, ``Black Hole Bound on the Number of Species and Quantum Gravity at LHC,''
Phys. Rev.  {\bf D77} ( 2008) 045027,  arXiv:0710.4344 [hep-th];
``Phenomenology of $10^{32}$ Dark Sectors", arXiv:0905.1709 [hep-ph]

R.~Brustein, G.~Dvali and G.~Veneziano, `` A Bound on Effective 
Gravitational Coupling from Semiclassical Black Holes",  
 JHEP 0910:085,2009,  arXiv:0907.5516 [hep-th]

\bibitem{zn}

G. Dvali , M. Redi, S. Sibiryakov,  A. Vainshtein, 
Gravity Cutoff in Theories with Large Discrete Symmetries, 
Phys.Rev.Lett.101:151603,2008,  arXiv:0804.0769 [hep-th].


\bibitem{gia-cesar} G.~Dvali and C~Gomez, ``Quantum Information and Gravity Cutoff in Theories with Species'',
Phys. Lett. {\bf B674}  (2009) 303,  arXiv:0812.1940 [hep-th] .

G.~Dvali and C.~ Gomez, 
 ``Strong Coupling Holography",  arXiv:0907.3237 [hep-th]

\bibitem{solod}
G. Dvali , S.N. Solodukhin, 
Black Hole Entropy and Gravity Cutoff,
 arXiv:0806.3976 [hep-th]

\bibitem{dvali-lust}

Evaporation of Microscopic Black Holes in String Theory and the Bound on Species.
G. Dvali , Dieter Lust,  arXiv:0912.3167 [hep-th]

\bibitem{gia-cesar2}
G.~Dvali and C.~Gomez,  ``Species and Strings",  arXiv:1004.3744. 

\bibitem{peskin}
 S. Cullen, M. Perelstein and M.E. Peskin,  TeV strings and collider probes of large extra dimensions, Phys. Rev. D 62, 055012 (2000) [arXiv:hep-ph/0001166].
 


\bibitem{dieter}

D. Lust , S. Stieberger, T.R. Taylor, 
 The LHC String Hunter's Companion,  Nucl.Phys.B808:1-52,2009, arXiv:0807.3333 [hep-th]

L.A. Anchordoqui, H. Goldberg , D. Lust , S. Nawata , S. Stieberger, T. R. Taylor,
Dijet signals for low mass strings at the LHC, 
Phys.Rev.Lett.101:241803,2008,  arXiv:0808.0497 [hep-ph]

L.A. Anchordoqui, H. Goldberg,  D. Lust, S. Nawata , S. Stieberger , T. R. Taylor, 
LHC Phenomenology for String Hunters,   Nucl.Phys.B821:181-196,2009, arXiv:0904.3547 [hep-ph]. 


D. Lust, O. Schlotterer, S. Stieberger, T.R. Taylor,
The LHC String Hunter's Companion (II): Five-Particle Amplitudes and Universal Properties.
 Nucl.Phys.B828:139-200,2010,  arXiv:0908.0409 [hep-th]


\bibitem{tao}
 
P. Burikham, T. Figy, T. Han,
TeV-scale string resonances at hadron colliders, 
 Phys.Rev.D71:016005,2005,  [hep-ph/0411094]

Z. Dong, T. Han, M. Huang, G. Shiu,  
Top Quarks as a Window to String Resonances, 
 arXiv:1004.5441 [hep-ph]

\bibitem{quantumcreation}

Grishchuk, L.P. and Zeldovich, Ya.B. (1982), Complete cosmological
theories, in Quantum Structure of Space and Time, ed. M. Duff and C.
Isham (Cambridge University Press, Cambridge); 

 Vilenkin, A. (1982). Creation of universes from nothing,
Phys. Lett. B117, 25. 

 Hartle, J.B. and Hawking, S.W. (1983),  Wave
function of the universe, Phys. Rev. D28, 2960;

  Linde, A.D,  Quantum creation of the inflationary universe, 
  Lett. Nuovo Cimento 39, (1984), 401.




\end{thebibliography}
\end{document}